\documentclass[aps,prd,a4paper,superscriptaddress,nofootinbib,10pt, two column]{revtex4-1}

\usepackage{amsmath,amssymb,graphicx,multirow,dcolumn,bm,latexsym,soul,nicefrac}
\usepackage{epstopdf}
\usepackage{mathrsfs}
\usepackage{acronym}
\usepackage[colorlinks,linkcolor=blue,citecolor=blue,urlcolor=blue ]{hyperref}
\usepackage{ulem}
\usepackage{subcaption}
\usepackage{float}
\usepackage{xcolor}
\usepackage{diagbox}

\usepackage[labelformat=simple,labelsep=period,skip=3pt]{caption}
\newcounter{RomanNumber}

\allowdisplaybreaks

\newcommand{\HBU}{Department of Physics, Hebei University, Baoding, 071002, China. \\
Hebei Key Laboratory of High-precision Computation and Application of Quantum Field Theory, Baoding, 071002, China. \\
Hebei Research Center of the Basic Discipline for Computational Physics, Baoding, 071002, China.}
\newcommand{\BIMSA}{Beijing Institute of Mathematical Sciences and Applications, Beijing 101408, China}
\newcommand{\SYU}{ School of Physics and Astronomy, Sun Yat-sen University (Zhuhai Campus), Zhuhai 519082, China. \\
MOE Key Laboratory of TianQin Mission, TianQin Research Center for
Gravitational Physics, Frontiers Science Center for TianQin, \\
 Gravitational Wave Research Center of CNSA, Sun Yat-sen University (Zhuhai Campus), Zhuhai 519082, China.}
\newcommand{\KIA}{Kavli Institute for Astronomy and Astrophysics at Peking University, 100871 Beijing, China}

\begin{document}

\title{Constraining modified theories of gravity through the detection of one extremely large mass-ratio inspiral}

\author{Hui-Min Fan}
\affiliation{\HBU}
\email{fanhm@hbu.edu.cn}

\author{Alejandro Torres-Orjuela}
\affiliation{\BIMSA}
\email{atorreso@bimsa.cn}

\author{Ver\'{o}nica V\'{a}zquez-Aceves}
\affiliation{\KIA}
\email{veronica@pku.edu.cn}

\author{Tian-Xiao Wang}
\affiliation{\SYU}

\author{Tai-Fu Feng}
\affiliation{\HBU}

\date{\today}

\begin{abstract}
Extremely large mass-ratio inspirals (XMRIs), formed by brown dwarfs inspiraling into a massive black hole, emit gravitational waves (GWs) that fall within the detection band of future space-borne detectors such as LISA, TianQin, and Taiji. Their detection will measure the astrophysical properties of the MBH in the center of our galaxy (SgrA$^\ast$) with unprecedented accuracy and provide a unique probe of gravity in the strong field regime. Here, we estimate the benefit of using the GWs from XMRIs to constrain the Chern-Simons theory. Our results show that XMRI signals radiated from the late stages of the evolution are particularly sensitive to differences between Chern-Simons theory and general relativity. For low-eccentricity sources, XMRIs can put bounds on the Chern-Simons parameter $\zeta$ at the level of $10^{-1}$ to an accuracy of $10^{-3}$. For high-eccentricity sources, XMRIs can put bounds on the parameter $\zeta$  at the level of $10^{-1}$ to an accuracy of $10^{-6}$. Furthermore, using the time-frequency MCMC method, we obtain the posterior distribution of XMRIs in the Chern-Simons theory. Our results show that almost all the parameters can be recovered within $1\sigma$ confidence interval. For most of the intrinsic parameters, the estimation accuracy reaches $10^{-3}$. For the brown dwarf mass, the estimation accuracy reaches $10^{-1}$, while for $\zeta$, the estimation accuracy reaches $\Delta\log_{10}\zeta=0.08$ for high eccentricity sources and 1.27 for low eccentricity sources.


\end{abstract}
\maketitle


\section{Introduction}
General relativity (GR) has provided accurate explanations for the majority of gravitational phenomena we observed~\cite{Psaltis:2008bb, Stairs:2003eg,Will:2014kxa}. However, there are still many unanswered questions that need to be resolved, such as the inadequacy of GR in describing primordial black holes, the lack of a unified theory combining GR and quantum field theory, and the initial singularity problem, among others~\cite{Yunes:2024lzm, Landsman:2021mjt}. In light of these limitations, various modified theories of gravity have been proposed to address these issues~\cite{Chamberlain:2017fjl,Yagi:2012vf, Canizares:2012is}. To verify these corrections to GR, current and upcoming detectors are crucial~\cite{EventHorizonTelescope:2019ths, LIGOScientific:2025kei, Kelley:2025yud}, as they will provide observations in both strong and weak gravity regimes. The methods include using the shadow images captured by the Event Horizon Telescope to constrain black hole parameters and deviations from the Kerr metric~\cite{Rodriguez:2024ijx, Meng:2023wgi}, as well as using light-travel time deviations to measure the graviton mass with Pulsar Timing Arrays~\cite{Wang:2023div}. With the direct detection of gravitational waves (GWs) by ground-based detectors~\cite{LIGOScientific:2016vbw}, testing theories of gravity using GWs has become an effective approach, providing a valuable complement to electromagnetic measurements.

Ground-based detectors are sensitive to high-frequency GWs generated by stellar-mass sources~\cite{KAGRA:2021duu, LIGOScientific:2020kqk}. Space-borne GW detectors, including LISA~\cite{Baker:2019nia,lisa_2024}, TianQin~\cite{TianQin:2015yph,Li:2024rnk}, and Taiji~\cite{Ruan:2018tsw}, designed with longer arm-lengths, are capable of detecting more massive sources radiating GWs in the mHz frequency band~\cite{lisa_2022a,lisa_2022b,tianqin_2021,tianqin_2024,taiji_2015,torres-orjuela_huang_2023,torres-orjuela_2024b}. Among their target sources~\cite{Liang:2021bde, Huang:2020rjf,Liu:2020eko, Wang:2019ryf, Shi:2019hqa, Korol:2017qcx, Ruan:2021fxq, Shuman:2021ruh, Klein:2022rbf}, one of the most important is extreme mass ratio inspirals (EMRIs)~\cite{Fan:2020zhy,Fan:2024nnp}. This type of source, characterized by a mass ratio of about $10^{-5}$, involves a stellar-mass compact object (CO) inspiraling into the central massive black hole (MBH) of galaxies, serving as a unique tool to test GR in the strong field regime~\cite{Amaro-Seoane:2007osp}. However, due to the back-reaction from the CO, EMRI waveforms in the strong field regime are complicated, requiring higher-order self-force calculations to solve the EMRI waveforms accurately~\cite{Wei:2025lva, Pound:2015tma, Burke:2023lno}. This will inevitably bring challenges to perform accurate data analysis for the testing of GR.

Compared with EMRIs, extremely-large mass ratio inspirals (XMRIs) involve brown dwarfs (BD) inspiraling into SMBHs~\cite{amaro-seoane_2019,vazquez-aceves_lin_2022,vazquez-aceves_lin_2024, Seoane:2025hlg}, forming GW sources with a very low mass ratio of $q \sim 10^{-8}$. Since the back-reaction depends on $q$, the CO's orbit closely follows a standard geodesic, meaning that approximations work better in calculating the orbit. Moreover, XMRIs exhibit higher event rates per galaxy as they stay much longer in band than EMRIs~\cite{amaro-seoane_2019}. Consequently, such systems occurring at the Galactic Centre (GC) during the mission lifetime of LISA, TianQin, and Taiji and being detected is plausible. Within this region, electromagnetic observations achieve better sensitivity and thus will enable comprehensive multi-messenger studies in the future~\cite{Rodriguez:2024ijx, Meng:2023wgi}. Recent estimates suggest that by the time the LISA and TianQin missions are operational, there will be about 15 eccentric and five circular XMRIs in our galactic center emitting GWs within the detectable range~\cite{vazquez-aceves_lin_2022,vazquez-aceves_lin_2024}. Using XMRIs to constrain modified theories of gravity thus becomes a feasible challenge. Here, we explore the possibility of using XMRIs to constrain the Chern-Simons (CS) theory~\cite{Yagi:2012vf, Canizares:2012is} as an example. 

The CS theory is constructed to test parity symmetry in gravitational physics, with the Einstein Hilbert action modified by adding a parity-violating Chern-Simons term~\cite{Yunes:2009hc,Cardoso:2009pk}. The orbital evolution of a binary in a post-Newtonian (PN) expansion in CS theory was derived in Ref.~\cite{Gair:2011ym, Vigeland:2011ji}; thus, it is straightforward to obtain the CS GW waveform, and one can focus on studying the usefulness of XMRIs in testing GR. Generally speaking, in this study, we aim to generate the XMRI waveforms in the CS theory using the method described in Ref.~\cite{Gair:2011ym}. With the fitting factor~\cite{Sun:2024nut, Chatziioannou:2017tdw, Mangiagli:2018kpu, Baird:2012cu}, we investigate the range of the coupling parameter $\zeta$, which controls the strength of the gravitational
deviation from GR, that can be distinguished. Furthermore, with the Fisher information matrix (FIM)~\cite{Vallisneri:2007ev, Rodriguez:2013mla}, we evaluate the estimation accuracy of $\zeta$ under different conditions. 

To realize the full scientific objectives of space-borne GW detectors with respect to XMRIs, one has to infer the distribution of the physical parameters of such sources from the data~\cite{Marsat:2020rtl}. The inference problem is related to producing samples from the posterior distribution for source parameters. In this study, we employ the time-frequency Markov Chain Monte Carlo (MCMC) method proposed for EMRIs~\cite{Strub:2025dfs} and, for the first time, we explore the posterior distribution of XMRI parameters with this method.

This paper is organized as follows. In Section II, we present the method to calculate the XMRI waveform and the deviations arising from the CS theory. In Section III, we present the data analysis methods we use in our study. In Section IV, we present our main findings, and in Section V, we give a summary and draw conclusions. Throughout this paper, we use natural units where $G=c=1$.


\section{XMRI waveforms}

To construct waveforms in modified theories of gravity, Yunes and Pretorius developed a systematic and model-independent approach known as the parameterized post-Einsteinian (ppE) framework~\cite{Yunes:2009ke, Yunes:2010qb, Cornish:2011ys}. They then took initial steps to extend this approach to EMRIs~\cite{Vigeland:2011ji}, but this extension only allowed for a subclass of metric deviations. Subsequently, Gair and Yunes developed the ppE approach to the Analytic Kludge (AK) method, constructing generic geodesics for the CO's orbit and deriving the EMRI waveform in CS theory~\cite{Gair:2011ym}. Here, we follow this approach to obtain the XMRI waveform. This choice is also consistent with the calculation method presented in Ref.~\cite{vazquez-aceves_lin_2022,vazquez-aceves_lin_2024}.


The AK waveform~\cite{Barack:2003fp} is calculated using the quadrupole formula, which depends on the CO's orbit. This orbit is characterized by the eccentricity $e$, the orbital frequency $\nu$, and the orientation of the orbit with three angles: $\lambda$ describing the inclination of the orbit relative to the SMBH spin, $\alpha$ representing the Lense-Thirring precession of the orbital angular momentum, and $\tilde{\gamma}$ which describes the orbital pericenter precession. Assuming the inclination angle $\lambda$ is constant, the evolution of the CO can be computed by integrating the ordinary differential equation~\cite{Barack:2003fp} for the other four quantities used in the AK waveform.

Compared to GR, the metric in dynamical CS gravity has an additional $(t, \phi)$ component,
\begin{equation}\label{equ:gcs}
\begin{aligned}
g^{\rm CS}_{t\phi}=&-\frac{2M^2 a r}{\rho^2}\sin^2\theta\\
&+\frac{5}{8}\zeta\frac{a M^5}{r^4}\left(1+\frac{12M}{7r}+\frac{27M^2}{10r^2}\right)\sin^2\theta,
\end{aligned}
\end{equation}
where $\rho^2=r^2+a^2M^2\cos^2\theta$, $a$ is the dimensionless MBH spin, $M$ is the MBH mass, and $\zeta$ is the dimensionless CS parameter that controls the strength of the gravitational deviation from GR and is related to $\xi$ by
\begin{equation}
\zeta=\xi/M^4,
\end{equation}
where $\xi$ is a universal parameter that falls off with distance as $r^{-4}$~\cite{Canizares:2012is}. According to the method in Ref.~\cite{Gair:2011ym}, the second term in Eq.~(\ref{equ:gcs}) results in a correction to the orbital evolution of the CO with
\begin{equation}\label{equ:CSPara}
\begin{aligned}
M \delta \dot{e}=&-5{\eta}{(2\pi M \nu)^{17/3}}(1-e^2)^{-7}\frac{\zeta a}{e \sin\theta}g_e,\\
2\pi M^2 \delta \dot{\nu}=&192 \eta (2\pi M\nu)^{20/3}{(1-e^2)^{-8}}\zeta a \sin\theta g_\nu,\\
M \delta \dot{\tilde{\gamma}}=&\frac{75}{64}(2\pi M\nu)^{4}(1-e^2)^{-9/2}a\zeta \sin\theta(8+12e^2+e^4),\\
M \delta \dot{\alpha}=&-\frac{5}{64}a\zeta(2\pi M\nu)^{4}(1-e^2)^{-9/2}(8+24e^2+3e^4),
\end{aligned}
\end{equation}
where $\eta:=\mu/M$ is the mass ratio, $\mu$ is the CO's mass, and $g_e$ and $g_\nu$ are functions of the eccentricity $e$, given explicitly in Ref.~\cite{Gair:2011ym}.

%

The orbit of the CO is obtained in CS theory by adding these corrections to the orbital evolution. From the orbit, the time evolution of the CO's inertial tensor around the MBH can be constructed. Projecting this quantity to the source direction ($\theta_S, \phi_S$) in ecliptic coordinates, we obtain the plus and cross polarizations of the GW. The two polarizations can be combined to form the complex amplitude
%
\begin{equation}
h(t)=h^+(t)-ih^\times(t).
\end{equation}
In this study, we perform our analysis considering detection by the space-based detector TianQin, attempting to explore the phase shift induced to an XMRI waveform by the CS parameter $\zeta$ compared to GR. We estimate the accuracy with which $\zeta$ can be determined as well as its effect on the estimation of other parameters. Furthermore, to directly illustrate the impact of the CS parameter on the phase mismatch of gravitational waveforms in Sec.~\ref{sec:Result}, we disregard the operation mode ``3 month on +3 month off" of TianQin~\cite{TianQin:2015yph}.


\section{Methods for data analysis}

\subsection{Fitting Factor}\label{subsec:FF}

TianQin is a geocentric space-borne detector~\cite{TianQin:2015yph}, aiming to detect GW sources in the frequency band $10^{-4}\sim1$\,Hz. The detector consists of three satellites, forming a regular triangular constellation with an arm length of $L=\sqrt{3}\times 10^8$\,m. Its orientation points to the reference source RX J0806.3+1527, and it is expected to operate for five years. The noise model of TianQin is encoded in the following sensitivity curve~\cite{tianqin_2021,torres-orjuela_huang_2023}
\begin{equation}\label{equ:TQSn}
\begin{aligned}
S_n(f) =& \frac{20}{3}\frac{1}{L^2}\left[\frac{4S_a}{(2\pi f)^4}\left(1+\frac{10^{-4}\rm Hz}{f}\right)+S_x\right] \\
&\times\left[1+0.6\left(\frac{f}{f_*}\right)^2\right]
\end{aligned}
\end{equation}
where $S^{1/2}_a=1\times 10^{-15} \rm ms^{-2}/Hz^{-1/2}$ is the residual acceleration noise, $S^{1/2}_x=1\times10^{-12} \rm m/Hz^{1/2}$ is the position noise, and $f_*=1/(2\pi L)$ is the transfer frequency. 

The additional parameter $\zeta$ in CS theory will cause a phase mismatch compared to the waveform in GR. By extracting this information embedded in the detected data, the modified theory can be verified. Here, we aim to explore the distinguishability based on this waveform mismatch. The method applied is the fitting factor~\cite{Sun:2024nut, Chatziioannou:2017tdw, Mangiagli:2018kpu, Baird:2012cu}, which quantifies the overlap of two different waveform models as
\begin{equation}
FF=\frac{(h(\Theta)|h^\prime(\Theta))}{\sqrt{(h(\Theta)|h(\Theta))(h^\prime(\Theta)|h^\prime(\Theta))}}
\end{equation}
where $h(\Theta)$ is the XMRI waveform in GR, $h^\prime(\Theta)$ is the corresponding waveform in modified CS theory, and $\Theta$ stands for all parameters of the source. The noise-weighted inner product between signals $s_1(t)$ and $s_2(t)$ can be written as
\begin{equation}\label{equ:inprod}
(s_1|s_2)=2\int^\infty_0\frac{\tilde{s}_1(f)\tilde{s}^*_2(f)+\tilde{s}_2(f)\tilde{s}^*_1(f)}{S_n(f)}df
\end{equation}
where $\tilde{s}(f)$ is the Fourier transform of $s(t)$ and $S_n(f)$ is the power spectral density of TianQin as expressed in Eq.~(\ref{equ:TQSn}).

The mismatch between the two signals can then be obtained as
\begin{equation}
\mathcal M=1-FF.
\end{equation}
The threshold value $\mathcal {M}_{\rm th}$ that serves as a criterion to evaluate whether the mismatch between the two waveforms can be distinguished is
\begin{equation}\label{equ:ffthresd}
\mathcal M_{\rm th}=\frac{D}{2\rho^2}
\end{equation}
where $D$ is the number of parameters considered in the waveform and $\rho$ is the signal-to-noise ratio (SNR) of the source. The SNR also determines if the source can be detected in the first place, where we adopt a threshold of $\rho=20$ from Ref.~\cite{vazquez-aceves_lin_2024}.

\subsection{Fisher Information Matrix}

To verify the modified theory, determining the source parameters is necessary. However, the presence of noise will introduce uncertainties in the inference of these parameters, potentially leading to inaccurate estimations. To quantify these uncertainties, the Fisher information matrix (FIM) method is often employed for quick exploration~\cite{Vallisneri:2007ev, Rodriguez:2013mla}. Although the obtained $\Sigma$ only represents the Cramer-Rao bound of the covariance matrix, it can serve as a guide for Bayesian inference, as Bayesian calculations are much slower and more computationally costly. The FIM matrix is defined as
\begin{equation}
\Gamma_{ij}=\left(\frac{\partial\tilde{h}(f)}{\partial\Theta^i}\bigg|\frac{\partial\tilde{h}(f)}{\partial\Theta^j}\right)
\end{equation}
with $\Theta^i$ the $i$'th parameter of the waveform model. The Cramer-Rao bound of the covariance matrix is then obtained as $\Sigma = \Gamma^{-1}$. 


From the covariance matrix, the uncertainty $\sigma_{\Theta^i}$ can be determined as
\begin{equation}\label{equ:SigmaTh}
\sigma_{\Theta^i}=\Sigma_{\Theta^i\Theta^i}^{1/2}.
\end{equation}
Here, we are particularly interested in the determination of the CS parameter $\zeta$, which corresponds to the expression
\begin{equation}\label{equ:SigmaZeta}
\sigma_\zeta=\Sigma_{\zeta\zeta}^{1/2}.
\end{equation}

\subsection{Bayesian Inference}

Based on the results from the FIM, we explore the Bayesian inference method to infer the XMRI parameters. The Bayesian framework to infer the posterior probability distribution of the source parameters has the expression
\begin{equation}
p(\Theta|D)=\frac{p(D|\Theta)\pi({\Theta})}{p(D)},
\end{equation}
where $\pi(\Theta)$ is the prior distribution of the parameters, $p(D|\Theta)$ is the likelihood of observing the data $D$ given the parameters $\Theta$, and $p(D)$ is the evidence, which can be regarded as a normalization factor and does not need to be considered during the parameter estimation analysis. For stationary Gaussian noise, the likelihood for the GW signal can be obtained using the following expression
\begin{equation}
\ln{\mathcal L}\propto -\frac{1}{2}(D-h(\Theta)|D-h(\Theta))
\end{equation}
where the inner product is defined in Eq.~(\ref{equ:inprod}).

Compared to EMRIs, the modes of XMRIs exhibit lower evolution. However, the non-local parameter degeneracies of the XMRI sources persist, and numerous local maxima are present within the parameter space. These degeneracies cause the MCMC chains to get stuck on local maxima, preventing the detection of the true global maximum. Here, we explore a method to overcome this problem and make a first attempt at performing a Bayesian analysis for XMRIs. Drawing from EMRI studies, methods to address these challenges include semi-coherent searches, phenomenological waveform searches, harmonic search methods, and the adoption of convolutional neural networks to identify source parameters~\cite{Ye:2023lok,Wang:2012xh,Gair:2004iv,Zhang:2022xuq}. These approaches attempt to infer the EMRI source parameters by constraining the prior to a region around the primary peak, and thus need further development. The recent method proposed by Ref.~\cite{Strub:2025dfs} employs a time-frequency MCMC approach to infer EMRI parameters and has demonstrated both a loose prior range and effective EMRI parameter extraction capabilities. Therefore, we explore the application of this method in inferring XMRI parameters within the CS theory in this study. 

This new search method adopts two channels to enhance the search for the true global maximum. The first channel is the short-time Fourier transform. It defines a new inner product expressed as follows
\begin{equation}
(x|y)_{\rm{tf},\lambda}=\Bigg[\sum^{N-1}_{\tau=0} 4 \mathcal{R} \int^{\infty}_{0}\Bigg(\frac{|\tilde{x}_\tau(f)||\tilde{y}_\tau(f)|}{S_n(f)}\Bigg)^n\Bigg]^{1/n},
\end{equation}
where $\tilde{x}_\tau(f)$ is the short Fourier transform of the time series $x(t)$, $N$ is the number of time segments, and $n$ is the generalized mean that is used to enhance the efficiency of the parameter optimization algorithm. This statistic disregards phase coherence but imposes frequency evolution coherence between pixels, increasing the primary mode range in the parameter search volume. The second channel is parallel tempering MCMC. Although the primary mode range has become larger, the samplers still need to transition from one peak to another. This involves undergoing a series of downward jumps with lower probability, which is challenging in MCMC. The parallel tempering MCMC addresses this by assigning the posterior probability with
\begin{equation}
p_{T}(\Theta|D)\propto p(D|\Theta)^{1/T}\pi(\Theta),
\end{equation}
where $T$ is the temperature that controls the flattening of the posterior probability, allowing the sampler to explore more broadly. Through communication between multiple chains, the true global maxima can be effectively identified. This operation can be achieved using \texttt{Eryn}~\cite{Karnesis:2023ras,michael_katz_2023_7705496}, which is an advanced version of \texttt{emcee}~\cite{foreman2013emcee}.

\section{Results}\label{sec:Result}

\subsection{Waveform Overlap}\label{subsec:FFResult}

The estimation in Ref.~\cite{vazquez-aceves_lin_2024} suggests that about 15 eccentric and five circular XMRIs could be detected by space-borne detectors. In this study, we select six sources, including three with low eccentricity and three with relatively high eccentricity, to perform the analysis. The chosen sources are listed in Table~\ref{tabel:sources}. The parameters in this table are the same as in Ref.~\cite{vazquez-aceves_lin_2024}, except that we separate the semi-latus rectum later to examine the influence of the CS parameter on XMRI waveforms at different stages of the evolution. In this table, the first column corresponds to the source name we have chosen, while the second, third, and fourth columns correspond to the spin, the inclination angle, and the eccentricity of the XMRIs, respectively.

\begin{table}[t]
\caption{Parameters of the chosen sources. The first column indicates the source's name chosen; the second, third, and fourth columns correspond to the spin, the inclination angle, and the eccentricity of the XMRIs, respectively.}
\begin{center}
\begin{tabular}{c|c|c|c}
\hline
Sources & Spin ($a$) & Inclination angle ($\lambda\,[\rm rad]$) & Eccentricity ($e$)\\
\hline
s1 &0.1& 0.7&0.258564\\
s2 &0.1&1.57&0.279065\\
s3 &0.9&0.7&0.280562\\
\hline
s4 &0.9&1.0&0.410115\\
s5 &0.1&0.1&0.506363\\
s6 &0.9&0.4&0.581592\\
\hline
\end{tabular}
\end{center}
\label{tabel:sources}
\end{table}

Compared to GR, XMRIs in CS theory have an additional parameter, $\zeta$, resulting in a total of 15 parameters needed to describe their waveform. For $\zeta$, it is evident that a larger value leads to a greater waveform mismatch, making it easier to distinguish CS theory from GR. Here, we aim to explore the observation bounds that space-borne detectors could put on $\zeta$ under different conditions. We consider $\log_{10}\zeta$ with a value range in $(-5.0, -1.0)$. This corresponds to a bound on $\xi^{1/4}$ at a level of about $3\cdot10^5$\,km to $3\cdot10^6$\,km that is three to two orders of magnitude better than current Solar System bounds~\cite{Canizares:2012is, Ali-Haimoud:2011zme}. The selected orbits for low-eccentricity XMRIs have semi-latus recti of $p=8.5, 10.5, 12.5 M$, while for relatively high-eccentricity XMRI sources, the semi-latus rectum values are $p=14.5, 16.5, 18.5 M$. The relatively high semi-latus recti chosen for the latter sources is to guarantee a sufficiently long evolution time, as high-eccentricity orbits evolve more rapidly and can plunge quickly into the MBH. In regard to TianQin's mission lifetime of five years, we consider $T=0.5, 1.0, 2.0, 5.0$ years for the evolution time of these sources.

\begin{figure*}
\centering
\includegraphics[width=0.82\textwidth,clip=true,angle=0,scale=1.1]{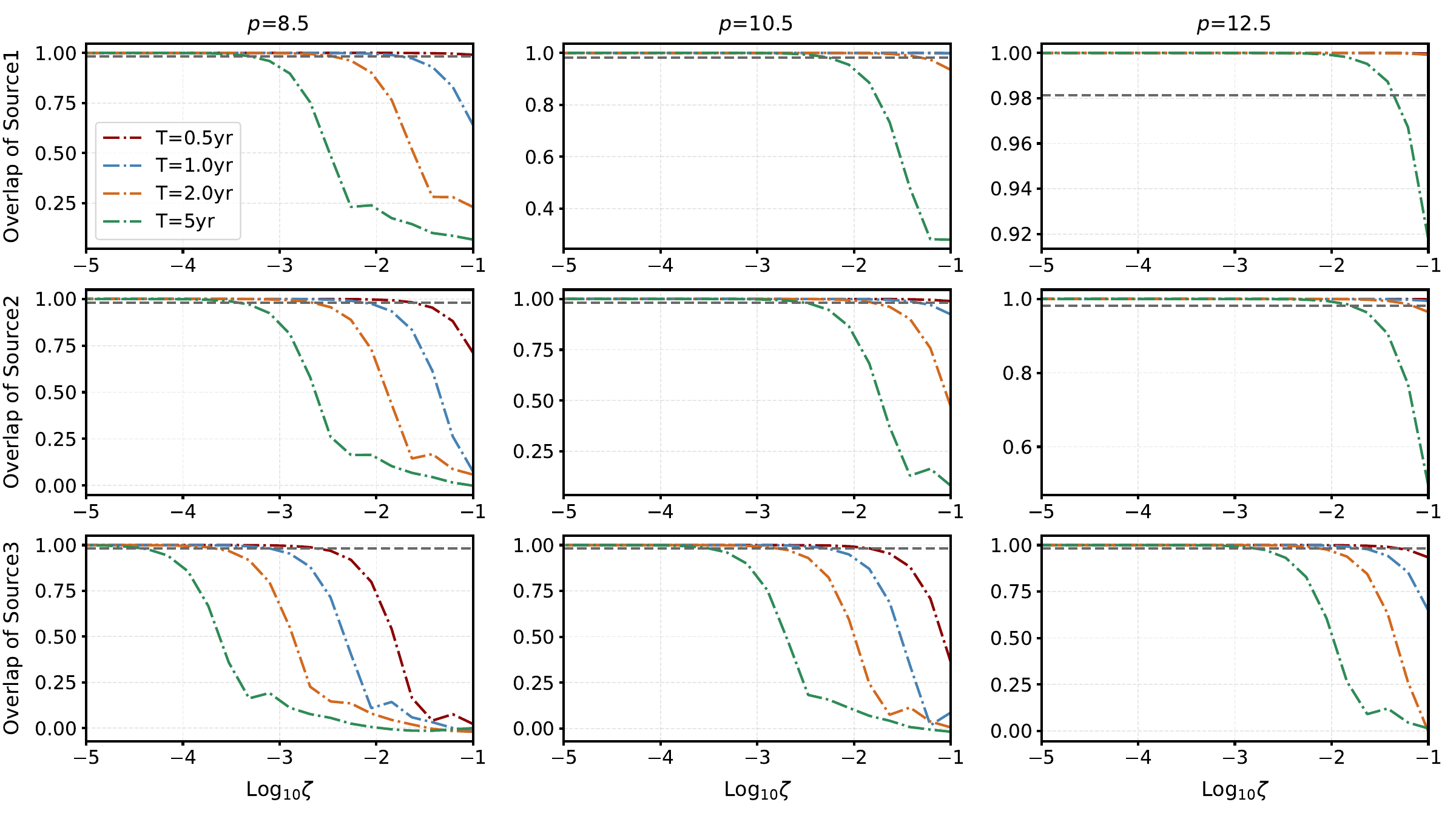}
\caption{The overlap results for s1, s2 and s3 from Table~\ref{tabel:sources}, where $p$ is given in units of mass $M$. The red, blue, orange, and green dash-dotted lines correspond to $T=0.5, 1.0, 2.0, 5.0$ years, respectively. The gray dashed lines correspond to the threshold value that can be distinguished. }
\label{fig:overlap1}
\end{figure*}
\begin{figure*}
\centering
\includegraphics[width=0.82\textwidth,clip=true,angle=0,scale=1.1]{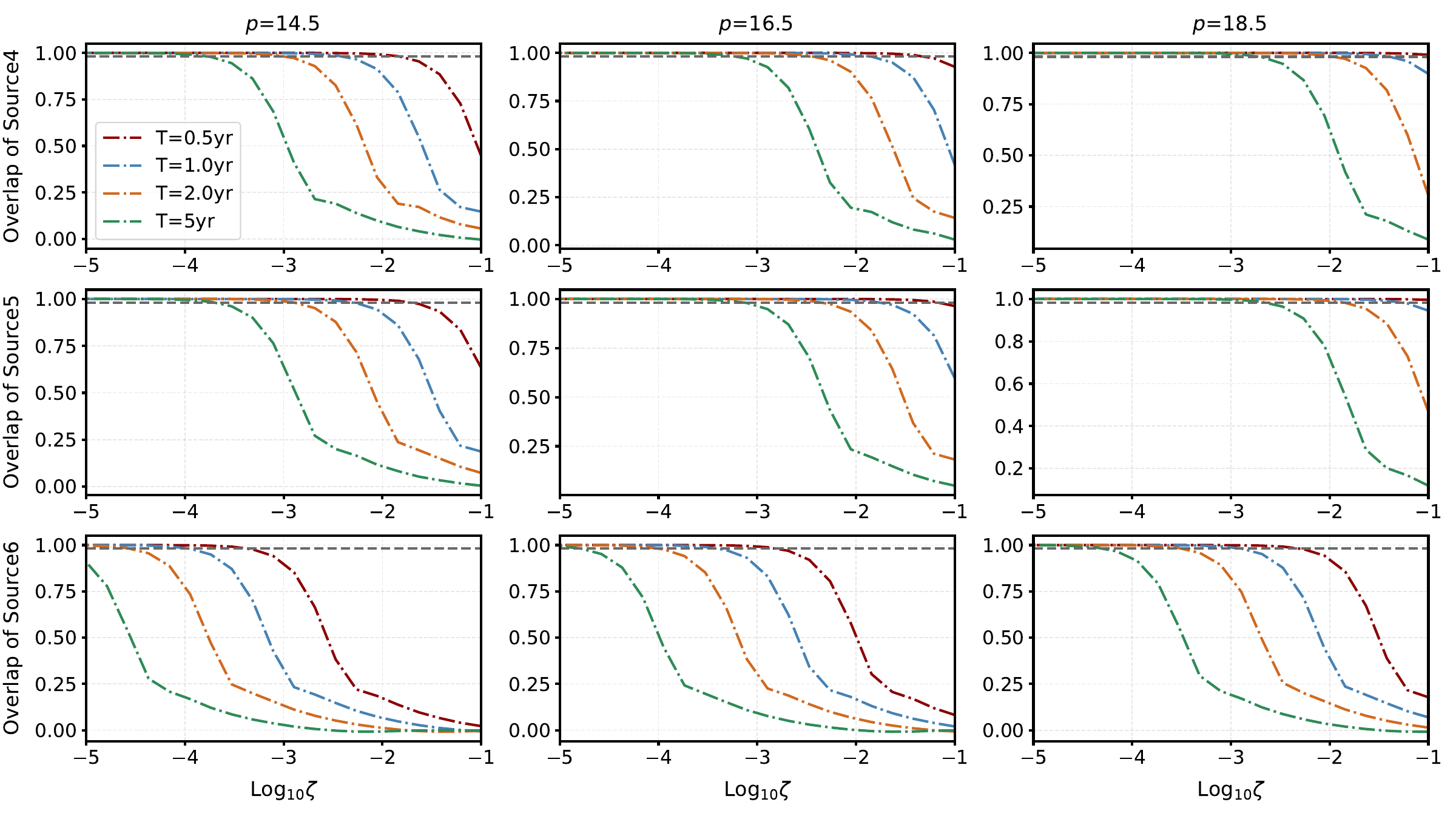}
\caption{The overlap results for sources s4, s5, and s6, respectively, where $p$ is given in units of mass $M$. We use the same convention as in Fig.~\ref{fig:overlap1}. }
\label{fig:overlap2}
\end{figure*}

In addition to the parameters mentioned above, there are ten more parameters that need to be determined in order to obtain the waveform. The detectable XMRIs are expected to occur within our galaxy. Thus, for the sky location, we choose $(\theta_S, \phi_S)$ equal to $(-0.096, 4.6574)\,[\rm rad]$, which corresponds to the direction of SgrA$^\ast$ in ecliptic coordinates. For the direction of the spin $a$ relative to the line of sight, we choose $(\theta_K, \phi_K)$ equal to $(1.0524, 1.6807)\,[\rm rad]$, along with the north pole of the Galactic disk. The mass $M$ of SgrA$^\ast$ and the BD mass $\mu$ are described as $4.0\cdot 10^{6} M_\odot$ and $0.05 M_\odot$, respectively, and the luminosity distance is given as $8.3$\,[\,kpc]~\cite{Gillessen:2008qv}. Finally, the three initial phase angles are randomly chosen with $\phi_0=1.0\,[\rm rad], \gamma_0=2.0\,[\rm rad], \alpha_0=3.0\,[\rm rad]$. The results for the overlap of the signals are displayed in Fig.~\ref{fig:overlap1} and Fig.~\ref{fig:overlap2}.

In these two figures, the x-axis corresponds to the values of $\log_{10}\zeta$, and the y-axis corresponds to the waveform overlap in CS theory and GR theory. The red, blue, orange, and green dash-dotted lines correspond to the overlap result with evolution times of $T=0.5, 1.0, 2.0, 5.0$ years, respectively. The gray dashed lines show the threshold value according to Eq.~(\ref{equ:ffthresd}) that can be distinguished. The first row in Fig.~\ref{fig:overlap1} corresponds to the overlap result of s1 in Table~\ref{tabel:sources} for orbits with $p=8.5, 10.5, 12.5\,M$, respectively. The second row corresponds to the overlap result of s2, and the third row corresponds to the overlap result of s3. Similarly, the first row in Fig.~\ref{fig:overlap2} shows the overlap results for s4 in Table~\ref{tabel:sources} for orbits with $p=14.5, 16.5, 18.5\,M$, respectively, while the second and third rows correspond to the overlap results of s5 and s6, respectively.

From Fig.~\ref{fig:overlap1}, we can observe that the longer the evolution time and the closer the BD is to SgrA$^\ast$, the easier it is to distinguish CS theory from GR. For s1, when $p$ equals $12.5\,M$, the waveforms are indistinguishable for $T=0.5, 1.0, 2.0$ years. Even when the waveform length reaches five years, the observations can only constrain the parameter to $\zeta=10^{-1.3}$. When $p$ equals $10.5\,M$, a waveform length of two years can put constraints to $\zeta=10^{-1.3}$. When the waveform length reaches five years, the bound improves to $\zeta=10^{-2.2}$. When $p$ equals $8.5 M$, which places the BD very close to SgrA$^\ast$, the bounds improve significantly. For a waveform length of $T=1.0$ year, it can put a constraint on $\zeta=10^{-1.7}$. For waveform lengths of $T=2.0$ and $T=5$ years, the bounds improve further to $\zeta=10^{-2.5}$ and $\zeta=10^{-3.5}$, respectively. 

The similar results from s2 and s3 show that for low eccentric XMRI sources, the region near the MBH is the best area for constraining the modified theory. For s2, a waveform length of $T=5$ years is still required to put bounds on the CS parameter below $10^{-1}$ for $p=12.5M$. However, for $p=8.5M$, a waveform length of $T=0.5$ years could constrain $\zeta$ to $10^{-1.5}$. Further information reveals that s3 performs relatively better in constraining CS theory. For $p=12.5M$, a waveform length of $T=2.0$ years could constrain $\zeta$ to $10^{-1.9}$, and a waveform length of $T=5$ years could constrain $\zeta$ to $10^{-2.8}$. For $p=8.5M$, a waveform length of $T=0.5$ years could constrain $\zeta$ to $10^{-2.3}$, and a waveform length of $T=5.0$ years could constrain $\zeta$ beyond $10^{-4.3}$. This is because in s3, the MBH has a high spin value. From Eq.~(\ref{equ:CSPara}), we can identify a direct relationship between the spin $a$ of the MBH and the resulting mismatch of the waveform.

From Fig.~\ref{fig:overlap2}, we can see that although the semi-latus rectum is larger for s4, s5, and s6, their ability to constrain CS theory is significantly better. For s6, when $p=18.5M$, a waveform length of $T=0.5$ years can constrain $\zeta$ to $10^{-2}$, and a waveform length of $T=5$ years can constrain $\zeta$ to $10^{-4.2}$. When $p=14.5M$, s6 can constrain $\zeta$ below $10^{-5}$ with a length of $T=5$ years. This strong performance for s6 is because it has a high eccentricity as well as a high spin of the MBH. From the explicit terms of $g_e$, $g_\nu$, and $a$ in Eq.~(\ref{equ:CSPara}), we can see a significant positive impact of these two parameters on the constraining ability of XMRI sources. The sources s4, s5 do not perform as well as s6, but they still perform better than s1, s2, s3. We thus can conclude that having a high spin makes constraining CS theory easier, but that having high eccentricity sources helps even more. 

Besides the above, we can observe a fluctuation of the lines around the overlap position of 0.25 in  Fig~\ref{fig:overlap1} and Fig.~\ref{fig:overlap2}. This occurs because the phase difference increases too much as the waveform overlap decreases, which may result in smaller $\Delta\phi$ during specific time intervals.

\subsection{Parameter Estimation of FIM}\label{subsec:ResultFIM}

To assess the estimation accuracy of $\zeta$ under different conditions, the FIM provides a quick and computationally efficient method.  In this section, we explore how accuracy of $\zeta$ can be determined and probe the ability of XMRIs to constrain the CS theory. 

Different from waveform overlap, which is independent of the SNR by dividing its contribution in Eq.~(\ref{equ:inprod}), there are two factors that affect the accuracy of parameter estimation. One is the SNR $\rho$, where the uncertainty $\sigma$ is generally inversely proportional to it. This can be seen by considering that if we scale the amplitude $h(t)$ by a factor $C$, making it $C\cdot h(t)$, the SNR will become $C\cdot \rho$, and the covariance matrix will become $1/C^2\cdot\Sigma$ according to  Eq.~(\ref{equ:SigmaTh}). The other factor is the induced phase mismatch when the parameter deviates from its true value. This is related to the overlap result in Sec.~\ref{subsec:FFResult}. Usually, a parameter that is sensitive to the waveform mismatch will have better estimation accuracy, as this indicates that $h(t)$ has a large gradient. Here, we first show the SNR of these XMRIs assuming GR, and then demonstrate the resulting estimation accuracy for $\zeta$ under different conditions. 

The SNR results for those XMRIs with an evolution time of $T=0.5$ years are shown in Table~\ref{tabel:snr}. In this table, the first column corresponds to the source names, while the second, third, and fourth columns correspond to the SNR values at different orbits with $p=8.5, 10.5, 12.5\,M$ and $p=14.5, 16.5, 18.5\,M$, respectively. From this table, we can observe a very high SNR for the sources with $p=8.5\,M$. This is because we place the BD into orbits very close to the MBH, where they move more rapidly and thus induce a significant boost in the SNR. Based on the comparison of the SNRs in this table and the overlap results shown in Fig.~\ref{fig:overlap1} and Fig.~\ref{fig:overlap2}, we provide an explanation of the performance of XMRIs on constraining the CS theory.

\begin{table}[t]
\caption{The SNR of XMRIs introduced in Table~\ref{tabel:sources} under different values of the semi-latus rectum $p$.}
\begin{center}
\begin{tabular}{c|c}
\hline
\diagbox{Sources}{$p\,[M]$} & ~~~~8.5 ~~~~~~ 10.5 ~~~~~ 12.5~~~~\\
\hline
s1 & 830.4 ~~~~ 298.1 ~~~  133.4 \\
s2 & 534.5 ~~~~ 192.3 ~~~~  84.2  \\
s3 & 807.2 ~~~~ 298.4 ~~~  133.7 \\
\hline
\diagbox{Sources}{$p\,[M]$}& ~~~~~14.5 ~~~~~ 16.5 ~~~~~ 18.5~~~~~~\\
\hline
s4 & 103.2 ~~~~~ 57.5 ~~~~~~\,35.8 \\
s5 & 262.4 ~~~~ 149.3 ~~~~~ 92.1 \\
s6 & 388.7 ~~~~ 223.8 ~~~\, 137.8 \\
\hline
\end{tabular}
\end{center}
\label{tabel:snr}
\end{table}

\begin{figure*}
\centering
\includegraphics[width=0.96\textwidth,clip=true,angle=0,scale=1.1]{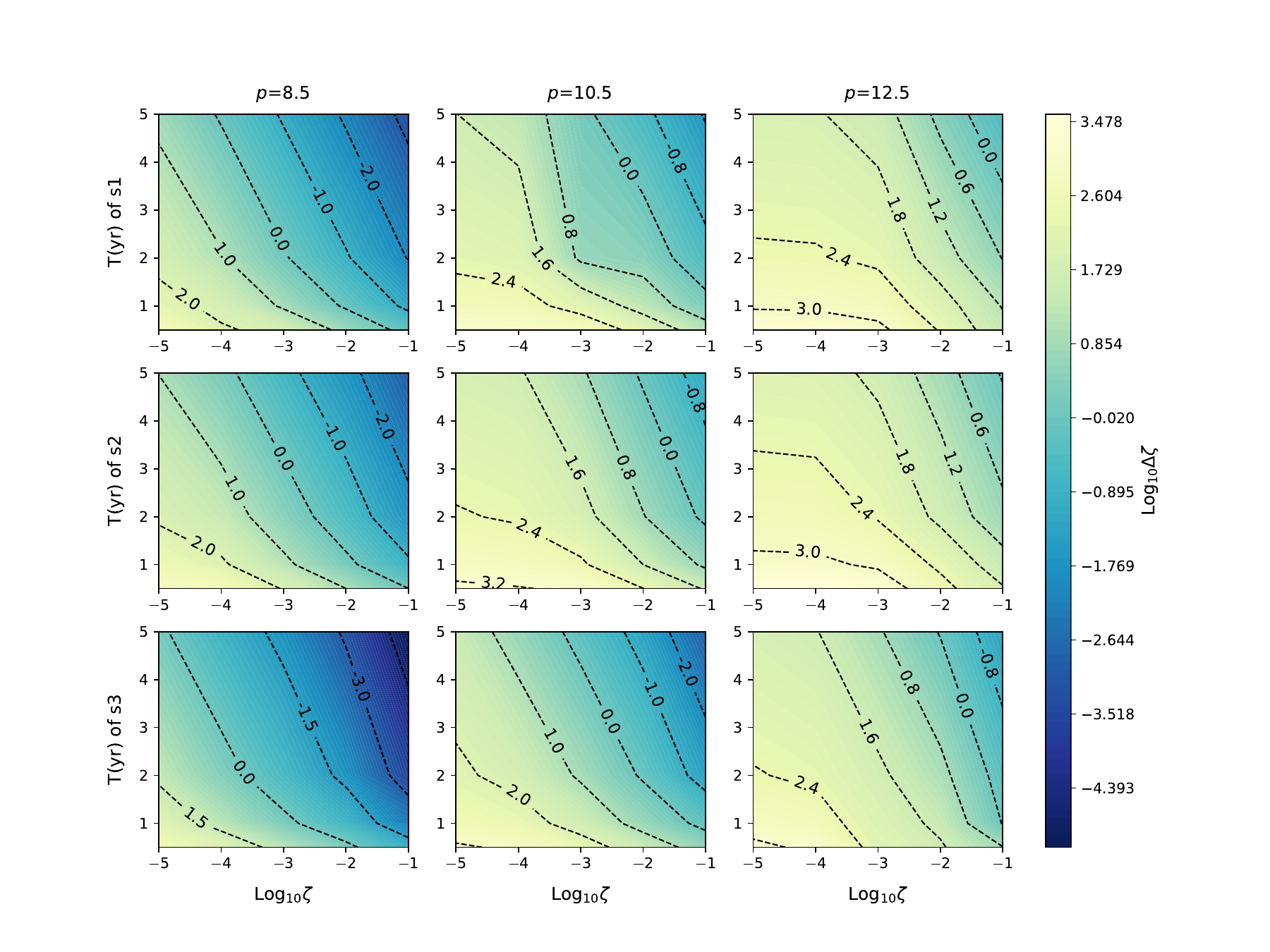}
\caption{The parameter estimation accuracy results of $\zeta$ for s1, s2 and s3, respectively, where the color indicates the estimation accuracy $\log_{10}\Delta\zeta$, the  x-axis represents the values of $\log_{10}\zeta$, and the y-axis represents the evolution time for different sources, $p$ is given in units of mass $M$.  }
\label{fig:ParaEsti}
\end{figure*}

\begin{figure*}
\centering
\includegraphics[width=0.96\textwidth,clip=true,angle=0,scale=1.1]{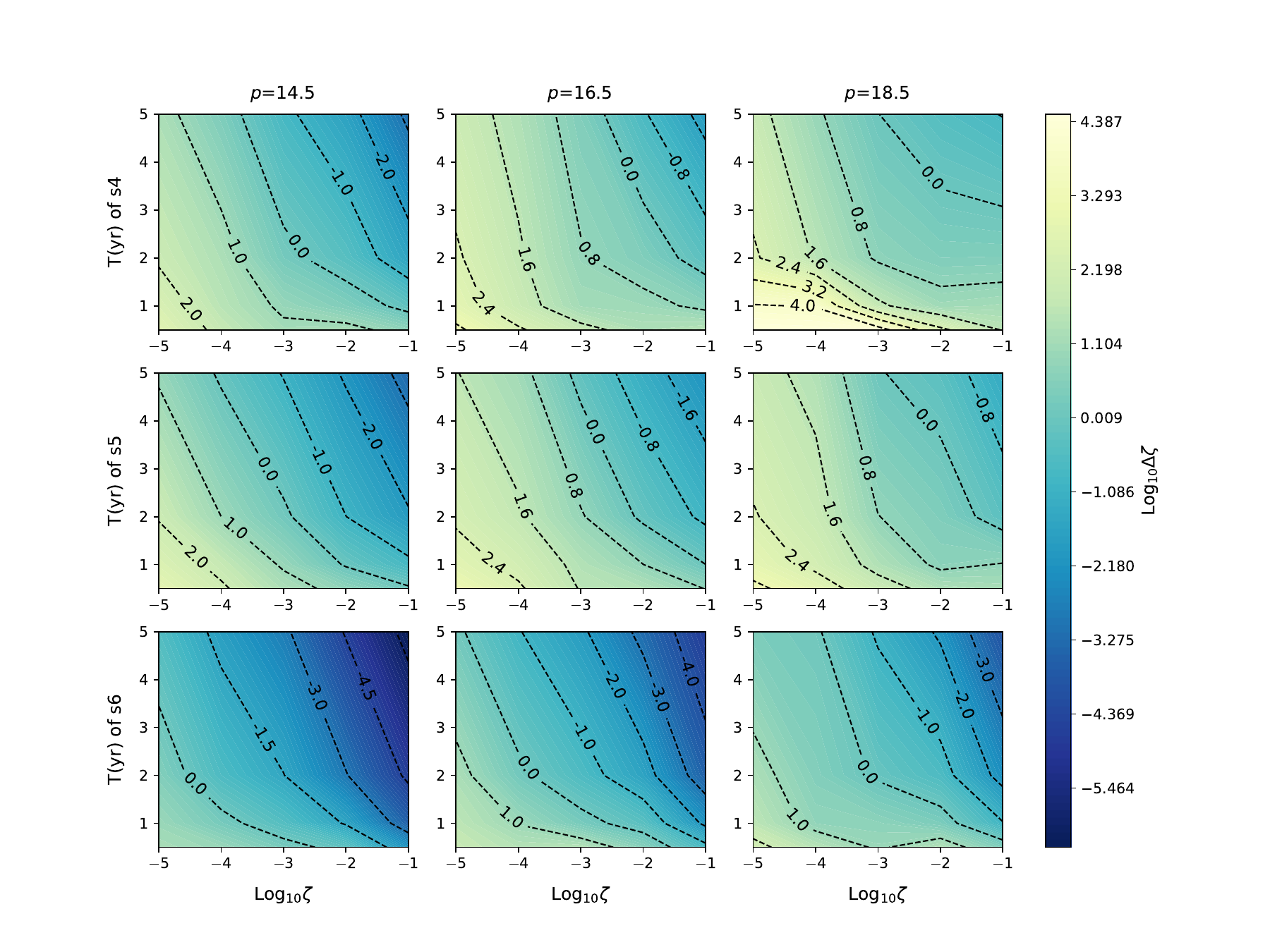}
\caption{The parameter estimation accuracy results of $\zeta$ for s4, s5 and s6, respectively, where $p$ is given in units of mass $M$. The same convention as in Fig.\ref{fig:ParaEsti}.}
\label{fig:ParaEsti2}
\end{figure*}

The results are shown in Fig.~\ref{fig:ParaEsti} and Fig.~\ref{fig:ParaEsti2}. In these two figures, the x-axis represents the values of $\log_{10}\zeta$, the y-axis represents the evolution time for different sources, and the color bars indicate the estimation accuracy of $\zeta$. The rows in the figures correspond to different sources, while the columns represent the sources for different initial semi-lati recti. Dark blue indicates better estimation accuracy compared to yellow. It is concentrated in the upper-right corner of these figures, corresponding to a longer evolution time and a larger $\zeta$.

The first and second rows of Fig.~\ref{fig:ParaEsti} show that the estimation accuracy of s1 is better than  s2 for $\zeta$. From Fig.~\ref{fig:overlap1}, we can see that the waveform mismatch of s2 is greater than that of s1 under the same conditions, while from Table~\ref{tabel:snr}, we see that the SNR of s2 is lower than that of s1. These observations highlight the positive influence of SNR on estimation accuracy. Meanwhile, when comparing s3 with s1, where the SNR is generally the same, the estimation accuracy for $\zeta$ is better for s3 due to the larger waveform mismatch. The same influence from SNR and the waveform mismatch is also evident in Fig.~\ref{fig:ParaEsti2}. When comparing the results from Fig.~\ref{fig:ParaEsti} and Fig.~\ref{fig:ParaEsti2}, we can find that although the SNRs for s4, s5, and s6 are generally smaller than those for s1, s2, and s3, the estimation accuracy for $\zeta$ is better, indicating the greater influence of phase mismatch on parameter estimation accuracy. 

Regarding the resulting accuracy for $\zeta$, the results for s1 indicate that when $p=12.5M$, a waveform length of $T=5$ years can constrain $\zeta=10^{-1}$ to an accuracy of $10^{-0.3}$. In contrast, for this orbit with a short waveform length and a small $\zeta$, the parameter estimation becomes very inaccurate, reaching up to $10^{3}$ for $T=1$ year and $\zeta=10^{-5}$. When $p=8.5M$, the results for s1 show that, a waveform length $T=1$ years could constrain $\zeta=10^{-1}$ to an accuracy of $10^{-1}$. Meanwhile, a waveform length of $T=5$ years is needed in order to constrain $\zeta=10^{-3}$ to an accuracy of $10^{-1}$. A similar demonstration for other sources is presented in Fig.~\ref{fig:ParaEsti} and Fig.~\ref{fig:ParaEsti2}.

Among all sources considered, s6 performs the best in constraining the CS parameter. This is due to its high SNR and large waveform mismatch. The results show that when $p=18.5M$, a waveform length of $T=5$ years could constrain $\zeta=10^{-1}$ to an accuracy of $10^{-3.3}$, and $\zeta=10^{-3}$ to an accuracy of $10^{-1}$. When $p=14.5M$, the results show that a waveform length of $T=4$ years could constrain $\zeta=10^{-1}$ to an accuracy $10^{-5}$ and $\zeta=10^{-5}$ to an accuracy of $10^{-0.5}$. Meanwhile, for a waveform length of $T=0.5$ year, the results shows that it could constrain $\zeta=10^{-1}$ to an accuracy of $10^{-3}$ and $\zeta=10^{-2}$ to an accuracy of $10^{-1}$. As most XMRI sources that can be detected have high eccentricity, this type of source will play an important role in constraining modified theories.

\subsection{The Bayesian Parameter Estimation Analysis}

Accurate parameter estimation is essential for analyzing the properties of XMRIs and for testing GR. Using the FIM method, we presented the estimation accuracy of $\zeta$ under different conditions in Sec.~\ref{subsec:ResultFIM}, providing a preliminary indication of the constraints on $\zeta$. Here, we aim to employ Bayesian inference to obtain the posterior distribution of the XMRI parameters. This approach will enable a more robust assessment of the potential for future detections. As a preliminary exploration, we take sources s1 and s6 as examples to examine the applicability of the time-frequency MCMC method~\cite{Strub:2025dfs} for XMRI parameter inference. To reduce computation cost, we adopt a short waveform length of $T=0.5$ years and choose a relatively large value of $\zeta=10^{-1}$ to ensure adequate estimation accuracy. No instrumental noise and response function are included in the data $D$.

In Stub's work~\cite{Strub:2025dfs}, by repeating the search several times first and selecting the set of parameters with the highest convolution power, the priors for those intrinsic parameters $(M, \mu, p_0, e_0)$ are narrowed and centered around the estimated values. In the present study, we skip this initial optimization and assume that such a narrowed range has already been obtained for those intrinsic parameters. The mass and the sky location ($\theta_S,\phi_S, D$) of SgrA$^\ast$ have been constrained with high accuracy through electromagnetic observations, which allow us to adopt narrow priors for these parameters~\cite{Gillessen:2008qv}. We, further, specify a flat prior distribution for the XMRI parameters, as listed in Table~\ref{tabel:Priors}. In this table, the first column corresponds to the XMRI parameters in CS theory, the second and third columns correspond to the input values and priors of s1, while the fourth and fifth columns correspond to the input values and priors of s6. From Table~\ref{tabel:Priors}, we can find a broader prior for s6 compared to s1. This is because the SNR of s6 is lower than that of s1, which leads to a wider posterior distribution for its parameters. A narrower prior would not sufficiently cover the full posterior distribution of s6.  


\begin{table*}[t]
\caption{Parameters, input values, and priors considered in the Bayesian parameter estimation for an XMRI for the examples sources s1 (second and third column) and s6 (fourth and fifth column).}
\begin{center}
\begin{tabular}{c|cc|cc}
\hline
Parameters & Input Values (s1) & Priors (s1) &  Input Values (s6) & Priors (s6)\\
\hline
$M\,[M_\odot]$&$4.0\cdot 10^6$&($3.995\cdot 10^6$, $4.005\cdot 10^6$)& $4.0\cdot 10^6$ &$(3.900\cdot10^6$, $4.100\cdot10^6)$\\
$\mu\,[M_\odot]$&0.05&(0.01, 0.20)& 0.05 &(0.01, 0.20)\\
$a$&0.1&(0.099, 0.101) & 0.9&(0.870, 0.930)\\
$e_0$&0.258467&(0.257467, 0.259467)& 0.581592 & (0.571592,0.591592)\\
$\theta_K\,[\rm rad]$&1.0524&(0.1, 2.1)& 1.0524& (0.1,  2.1)\\
$\phi_K\,[\rm rad]$&1.6807&(0.5, 3.0)& 1.6807& (0.1, 4.0)\\
$\theta_S\,[\rm rad]$&4.6574&(3.6, 6.0)& 4.6574&(2.0, 6.0)\\
$\lambda\,[\rm rad]$&0.7&(0.6, 0.8)& 0.4 &(0.2, 0.6)\\
$D\,(\rm kpc)$&8.3&(3.0, 22.0)& 8.3&(3.0, 22.0)\\
$p_0\,[M]$&8.50&(8.49, 8.51)& 14.5& (14.49, 14.51)\\
$\phi_0\,[\rm rad]$&1.0&(0.1, 2.0)& 1.0 & (0.1, 2.0)\\
$\gamma_0\,[\rm rad]$&2.0&(1.5, 2.5)& 2.0& (1.5, 2.5)\\
$\alpha_0\,[\rm rad]$&3.0&(2.5, 3.5)& 3.0& (2.5, 3.5)\\
$\log_{10}\zeta$&-1.0&(-5.0, 0.0)&-1.0&(-5.0, 0.0)\\
\hline
\end{tabular}
\end{center}
\label{tabel:Priors}
\end{table*}

\begin{figure*}
\centering
\includegraphics[width=0.96\textwidth,clip=true,angle=0,scale=1.1]{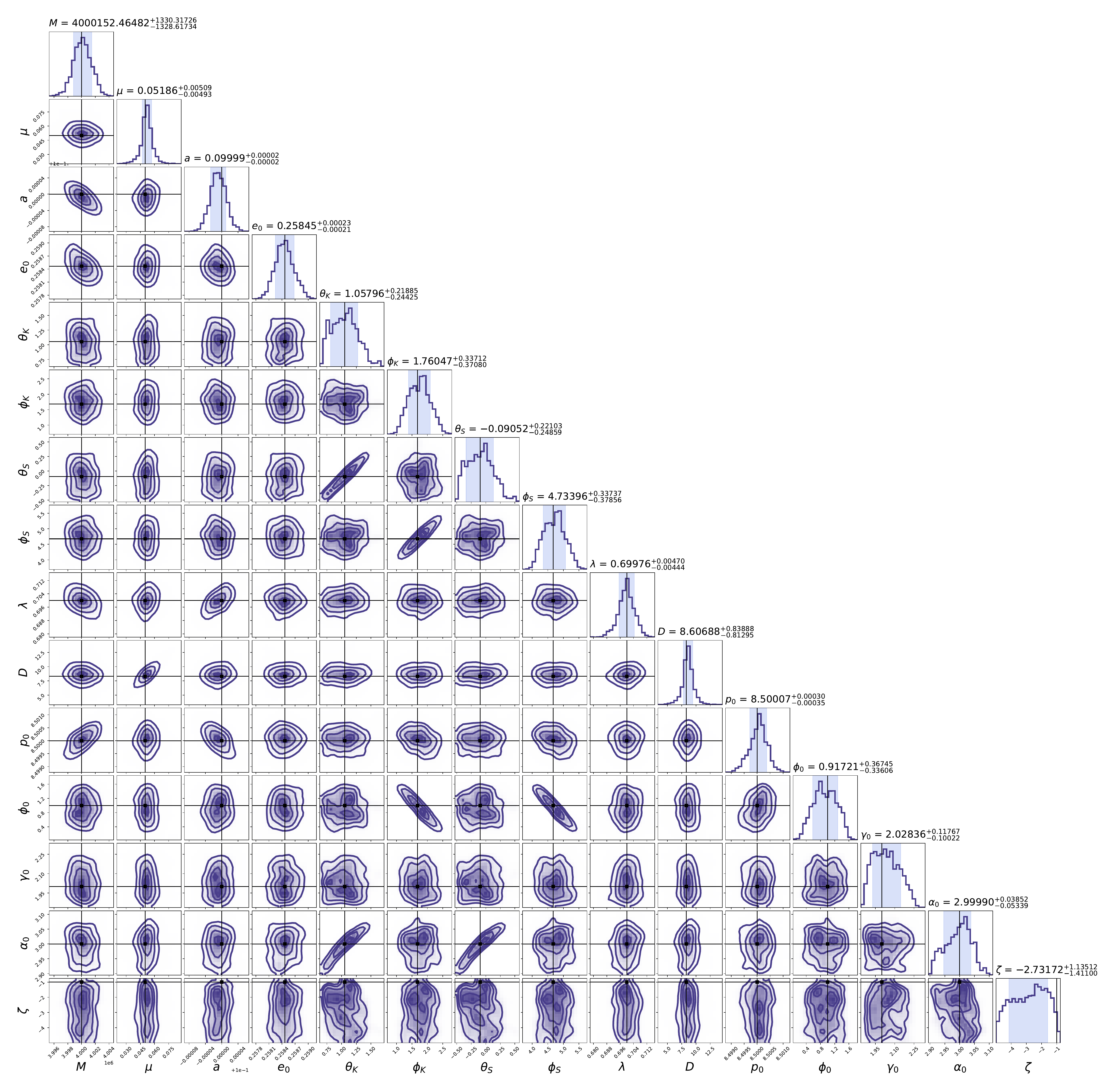}
\caption{Parameter estimation results using the time-frequency MCMC method for an XMRI with parameters as s1. Units: $M\,[M_\odot]$, $\mu\,[M_\odot]$, $\theta_K$\,[rad], $\phi_K$\,[rad], $\theta_S$\,[rad], $\phi_S$\,[rad], $\lambda$\,[rad], $D$\,[kpc], $p_0\,[M]$, $\phi_0$\,[rad], $\gamma_0$\,[rad], $\alpha_0$\,[rad]. }
\label{fig:BayParaEsti_s1}
\end{figure*}

\begin{figure*}
\centering
\includegraphics[width=0.96\textwidth,clip=true,angle=0,scale=1.1]{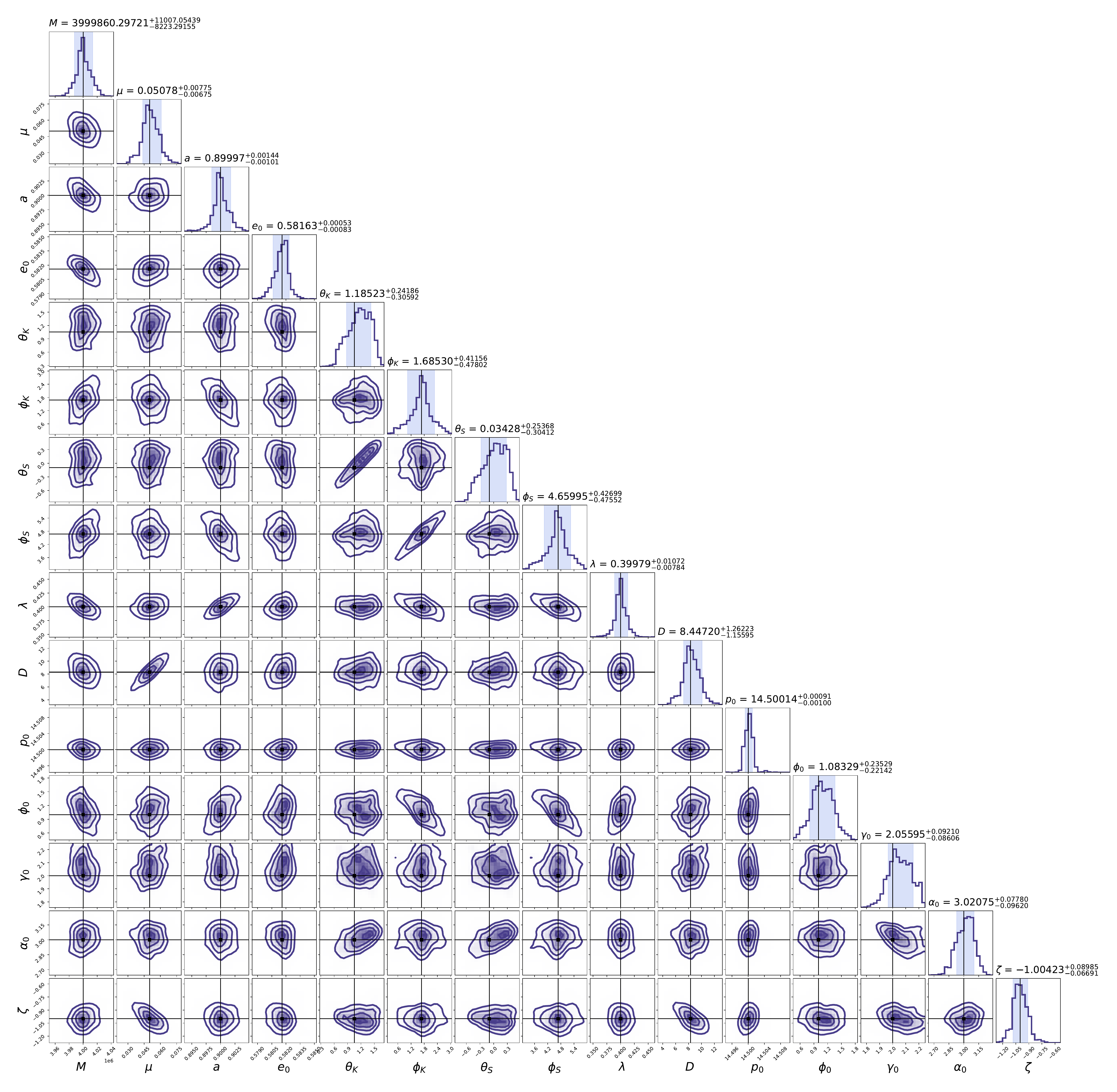}
\caption{Parameter estimation results using the time-frequency MCMC method for an XMRI with parameters as s6. Units: $M\,[M_\odot]$, $\mu\,[M_\odot]$, $\theta_K$\,[rad], $\phi_K$\,[rad], $\theta_S$\,[rad], $\phi_S$\,[rad], $\lambda$\,[rad], $D$\,[kpc], $p_0\,[M]$, $\phi_0$\,[rad], $\gamma_0$\,[rad], $\alpha_0$\,[rad].}
\label{fig:BayParaEsti_s2}
\end{figure*}

The constraint results for the XMRI parameters are summarized in Fig.~\ref{fig:BayParaEsti_s1} and Fig.~\ref{fig:BayParaEsti_s2} for s1 and s6, respectively. In these two figures, the blue contours represent the posterior distribution of the parameters, the black lines indicate the true values, the shadow regions denote the $1\sigma$ confidence interval, and the titles indicate the limits of the shadowed region. From these two figures, we see that the XMRI parameters can be well determined using the time-frequency MCMC method. The posterior peaks generally lie near the input values, and almost all parameters can be recovered within $1\sigma$ confidence interval.

From Fig.~\ref{fig:BayParaEsti_s1}, we can observe that, with the GW signal of s1, the parameters such as the mass of SgrA$^\ast$, the mass of the BD, the spin, the eccentricity, the inclination angle, and the initial semi-latum rectum have corresponding limits of $M=4000152^{+1330}_{-1328}\,[M_\odot]$, $\mu=0.05186^{+5.1\times10^{-3}}_{-4.9\times10^{-3}}\,[M_\odot]$, $a=0.09999^{+2.0\times10^{-5}}_{-2.0\times10^{-5}}$, $e_0=0.25845^{+2.3\times10^{-4}}_{-2.1\times10^{-4}}$, $\lambda=0.69976^{+4.7\times10^{-3}}_{-4.4\times10^{-3}}\,[\rm rad]$, and $p_0=8.50007^{+3.0\times10^{-4}}_{-3.5\times10^{-4}}\,[M]$, respectively. The estimation accuracies for these parameters are $\Delta M/M=3.3\times10^{-4}$, $\Delta \mu/\mu=0.097$, $\Delta a/a=2.0\times10^{-4}$, $\Delta e_0/e_0=8.5\times10^{-4}$, $\Delta\lambda=0.26\,\rm deg$, and $\Delta p_0/p_0=3.8\times10^{-5}$. Meanwhile, the spin direction of SgrA$^\ast$, the sky location, the luminosity distance, and the three initial phase angles have the corresponding limits of $\theta_K=1.05796^{+0.21885}_{-0.24425}\,[\rm rad]$, $\phi_K=1.76047^{+0.33712}_{-0.37080}\,[\rm rad]$, $\theta_S=-0.09052^{+0.22103}_{-0.24859}\,[\rm rad]$, $\phi_S=4.73396^{+0.33737}_{-0.37856}\,[\rm rad]$, $D=8.60688^{+0.83888}_{-0.81295}\,[\rm kpc]$, $\phi_0=0.91721^{+0.36745}_{-0.33606}\,[\rm rad]$, $\gamma_0=2.02836^{+0.11767}_{-0.10022}\,[\rm rad]$, and $\alpha_0=2.99990^{+0.03852}_{-0.05339}\,[\rm rad]$, respectively. The estimation accuracies for these parameters are $\Delta\theta_K=13.27\,[\rm deg]$, $\Delta\phi_K=20.28\,[\rm deg]$, $\Delta\theta_S=13.45\,[\rm deg]$, $\Delta\phi_S=20.51\,[\rm deg]$, $\Delta D/D=0.096$, $\Delta\phi_0=20.15\,[\rm deg]$, $\Delta\gamma_0=6.24\,[\rm deg]$, and $\Delta\alpha_0=2.63\,[\rm deg]$. For the CS parameter $\zeta$, the corresponding limit is $\log_{10}\zeta=-2.73172^{+1.13512}_{-1.41100}$, which has a relatively low accuracy of $\Delta\log_{10}\lambda=1.27306$. 

Compared to s1, the parameters of s6 exhibit much wider posterior distributions, as illustrated in Fig.~\ref{fig:BayParaEsti_s2}. From this figure, we can find that the mass of SgrA$^\ast$, the mass of the BD, the spin, the eccentricity, the inclination angle, and the initial semi-latum rectum have the corresponding limits of $M=3999860^{+11007}_{-8223}\,[M_\odot]$, $\mu=0.0508^{+7.7\times10^{-3}}_{-6.7\times10^{-3}}\,[M_\odot]$, $a=0.9000^{+1.4\times10^{-3}}_{-1.0\times10^{-3}}$, $e_0=0.5816^{+5.0\times10^{-4}}_{-8.0\times10^{-4}}$, $\lambda=0.3998^{+1.1\times10^{-2}}_{-7.8\times10^{-3}}\,[\rm rad]$, and $p_0=14.5001^{+9.0\times10^{-4}}_{-1.0\times10^{-3}}\,[M]$, respectively. Their estimation accuracies are $\Delta M/M=2.4\cdot10^{-3}$, $\Delta \mu/\mu=0.14$, $\Delta a/a=1.3\cdot10^{-3}$, $\Delta e_0/e_0=1.1\cdot10^{-3}$, $\Delta\lambda=0.53\,[\rm deg]$ and $\Delta p_0/p_0=6.6\cdot10^{-5}$. At the same time, the spin direction of SgrA$^\ast$, the sky location, the luminosity distance, and the three initial phase angles have the corresponding limits of $\theta_K=1.1852^{+0.2419}_{-0.3059}\,[\rm rad]$, $\phi_K=1.6853^{+0.4116}_{-0.4780}\,[\rm rad]$, $\theta_S=-0.0343^{+0.2537}_{-0.3041}\,[\rm rad]$, $\phi_S=4.6599^{+0.4270}_{-0.4755}\,[\rm rad]$, $D=8.4472^{+1.2622}_{-1.1560}\,[\rm kpc]$, $\phi_0=1.0833^{+0.2353}_{-0.2214}\,[\rm rad]$, $\gamma_0=2.0559^{+0.0921}_{-0.0861}\,[\rm rad]$, and $\alpha_0=3.0207^{+0.0778}_{-0.0962}\,[\rm rad]$, respectively. Their estimation accuracies are $\Delta\theta_K=15.69\,[\rm deg]$, $\Delta\phi_K=25.48\,[\rm deg]$, $\Delta\theta_S=15.98\,[\rm deg]$, $\Delta\phi_S=25.85\,[\rm deg]$, $\Delta D/D=0.14$, $\Delta\phi_0=13.08\,[\rm deg]$, $\Delta\gamma_0=5.11\,[\rm deg]$, and $\Delta\alpha_0=4.98\,[\rm deg]$. For the CS parameter $\zeta$, the corresponding limit is $\log_{10}\zeta=-1.0042^{+0.0898}_{-0.0669}$. Its estimation accuracy,  $\Delta\log_{10}\lambda=0.08$, is better than that of s1, which is consistent with the results presented in Sec.~\ref{subsec:ResultFIM}.

From the above results, it is evident that the mass of SgrA$^\ast$, the spin, the eccentricity, the inclination angle, and the initial semi-latus rectum can be determined with high accuracy. This is because these intrinsic parameters contribute to the phase of the GWs, and a slight change can accumulate into large differences over the huge number of waveform cycles. In contrast, parameters such as the luminosity distance, the spin orientation, and the sky location affect only the amplitude of the GWs. Consequently, their estimation accuracies do not achieve the same level of accuracy as the intrinsic parameters. The three initial phases of $\phi_0, \alpha_0, \gamma_0$ will cause a phase shift, but do not accumulate phase difference over time, and thus are not well constrained. Among the intrinsic parameters, the mass of the BD and the CS parameter are the least well constrained, especially $\zeta$. This is because these two parameters have a very small impact on the waveform phase, and the induced phase mismatch does not accumulate sufficiently within the evolution time of $T=0.5$ years. 

Comparing Fig.~\ref{fig:BayParaEsti_s1} and Fig.~\ref{fig:BayParaEsti_s2}, we find that for the intrinsic parameters $(M, a, e_0, \lambda, p_0)$, the estimation accuracy of s1 is significantly better than that of s6 by several times. This is because these parameters of s1 produce a GW signal with a stronger GW amplitude and features that the TianQin detector is more sensitive to. Thus, they can be determined more accurately during data analysis. In contrast, parameters such as $(\mu, D, \theta_S, \phi_S, \theta_K, \phi_K)$ do not show a significant improvement in accuracy for s1 compared to s6. This is due to the strong degeneracies between pairs $(\mu, D)$, $(\theta_S, \theta_K)$, and $(\phi_S, \phi_K)$, which broaden the confidence intervals and reduce the estimation precision. Regarding $\zeta$, the much higher spin and eccentricity values of s6 enhance the waveform mismatch induced by this parameter relative to s1, thus leading to better estimation accuracy.


With the above estimation results, the XMRI parameters can be constrained to a precise range, enabling a better understanding of the astrophysical properties of this type of source, such as its host environment and the spacetime geometry of the Galaxy. Regarding GR testing, $\zeta=10^{-1}$ can be recovered from the posterior of s6 with an evolution time $T=0.5$ year. However, for s1, a length of $T=0.5$ years is not sufficient to accurately recover the value of $\zeta=10^{-1}$.  Given that TianQin has a mission lifetime of 5 years, the bounds put on $\zeta=10^{-2}$ for s1 and on $\zeta<10^{-5}$ for s6 can be expected, according to Fig.~\ref{fig:ParaEsti} and Fig.~\ref{fig:ParaEsti2}. Therefore, we expect that $\zeta=10^{-1}$ will be well recovered with a longer waveform length for s1.

%


\section{Summary and Conlusions}\label{conclusion}

Due to their extremely large mass ratios and proximity to us, XMRIs may relax the requirements on self-force calculations in testing GR, and serve as effective method that provide cross-validation for electromagnetic observations. In this study, we investigate the potential of XMRIs for constraining modified theories of gravity. We take CS theory as an example and utilize the AK model to generate the XMRI waveforms. We then employ the fitting factor to assess the distinguishable of CS theory from GR under different conditions. After that, we use the Fisher information matrix to estimate the accuracy of the CS parameter $\zeta$. Finally, we explore the posterior distribution of XMRI parameters using the time-frequency MCMC method.  

Our results show that for XMRIs, a BD that stays in an orbit close to SgrA$^\ast$ has a significant advantage in testing GR. For low eccentricity sources, a waveform length of $T=1$ year can distinguish CS theory from GR for $\zeta$ down to $10^{-1}$. However, if the semi-latus rectum is larger than $p=10.5M$, an extended observation period of $T=5$ years will be required to constrain the $\zeta$ to $10^{-1}$. Compared to the low-eccentricity XMRIs, high-eccentricity XMRIs perform better in constraining CS theory. For a high-eccentricity source (s6), a waveform length of $T=0.5$ can constrain $\zeta$ to $10^{-3}$. This enhances the potential of using XMRIs to test GR, as most detectable XMRI sources are expected to have high eccentricities~\cite{vazquez-aceves_lin_2022,vazquez-aceves_lin_2024}. The FIM results are closely related to both the SNR of the source and the waveform mismatch. In general, a higher SNR and a greater waveform mismatch lead to better estimation accuracy for $\zeta$. For a low-eccentricity source (s1), the estimation accuracy can reach $10^{-3}$ for a waveform length of $T=5$ years, when $p=8.5M$ and $\zeta=10^{-1}$. For a high-eccentricity source (s6), the estimation accuracy can reach $10^{-6}$ under the same conditions, when $p=14.5M$. Regarding the posterior distribution of sources s1 and s6, the results show that the XMRI parameters can be well determined using the time-frequency MCMC method. The posterior peaks generally lie near the injected values, and almost all parameters can be recovered within the $1\sigma$ confidence interval. 


\section*{Acknowledgements}
We are grateful to Shuo Sun and Chang-Qing Ye for their helpful discussion and advice. This work has been supported by Hebei Natural Science Foundation with Grant No. A2023201041, Postdoctoral Fellowship Program of CPSF under Grant Number GZC20240366. VVA acknowledges support from the PKU Postdoctoral Boya Fellowship. ATO acknowledges support from the Key Laboratory of TianQin Project (Sun Yat-sen University), Ministry of Education (China).  
~\\









\bibliographystyle{apsrev4-1.bst}
\bibliography{reference}


\end{document}